\documentclass[twocolumn,showpacs,preprintnumbers,amsmath,amssymb]{revtex4}
\usepackage{graphicx}

\begin{document}
\title{{\Large  New magic number for neutron rich Sn isotopes}}

\author{S. Sarkar
\thanks{e-mail: ss@physics.becs.ac.in}}
\affiliation{Department   of   Physics,  Bengal  Engineering  and
Science University, Shibpur, Howrah - 711103, INDIA}

\author{M. Saha Sarkar
\thanks{e-mail: maitrayee.sahasarkar@saha.ac.in}}
\affiliation{Nuclear  and Atomic Physics Division, Saha Institute
of Nuclear Physics, Kolkata 700064, INDIA }

\date{\today}

\begin{abstract}
The variation of $E(2^+_1)$ of  $^{134-140}Sn$ calculated with empirical SMPN interaction has striking similarity with that of  experimental $E(2^+_1)$ of even-even $^{18-22}O$ and $^{42-48}Ca$, showing clearly that N=84-88 spectra exhibit the effect of gradual filling up of $\nu(2f_{7/2})$  orbital which finally culminates in a new shell closure at N=90. Realistic two-body interaction CWG  does not show this feature. Spin-tensor decomposition of SMPN and CWG interactions and variation of their  components with valence neutron number reveals that the origin of the shell closure at $^{140}Sn$lies in the three body effects. Calculations with CWG3, which is obtained  by including a simple three-body monopole term in the CWG interaction,  predict decreasing $E(2^+_1)$ for  $^{134-138}Sn$ and a shell closure  at $^{140}Sn$. 

\end{abstract}

\maketitle

The evolution of shell structure away from stability  \cite{sorlin:1} has been a topic of intense theoretical \cite{otsuka:1,doba:1} and experimental \cite{ange:1,gor:1,schi:1} studies since last decade. Theoretical studies have identified different reasons for the phenomenon of shell evolution  with neutron or proton excess. Among the various
components of the nucleon - nucleon interaction, the spin-orbit, tensor part and three - body effect play  important roles in the shell evolutions \cite{sorlin:1}. Due to tensor interactions, nuclear mean field undergoes variations with neutron excess. This leads to  monopole migration \cite{otsuka:1}.  It is observed for both proton-rich as well as neutron-rich nuclei. While approaching the neutron drip line, the neutron density becomes very diffused \cite{doba:1} which can also lead to shell quenching. For exotic light nuclei the well established magic numbers for the stable nuclei are found to be modified or new magic numbers have evolved.  At least four doubly  magic oxygen isotopes have been observed \cite{nndc,gor:1,brwn:1}. They are $^{14}O$, $^{16}O$, $^{22}O$ and $^{24}O$. Brown and Richter \cite{brwn:1} have framed a generalised new rule for magic numbers, valid specially for lighter nuclei. For heavier nuclei, experimental production of neutron rich isotopes are more difficult. There are severe limitations in acquiring spectroscopic information on them due to their low production rates and lifetimes. However, the shell evolution for neutron -rich nuclei above the doubly magic $^{132}Sn$ core is recently a topic of great interest.  The Sn isotopes in particular, pose many interesting problems \cite{karta:1,brwn:2,sah:1,sah:2,cor:1,sah:3,fission09,schi:1} in the study of evolution of nuclear structure with increasing neutron number.The near constancy of the first $2^+$ energy of Sn isotopes for A=102 to 130 at $\simeq$ 1.2 MeV is a text book example for seniority conserved spectra \cite{casten}. But the two valence neutron isotope of Sn just above $^{132}Sn$, {\it i.e.}, $^{134}Sn$, shows a sudden depression in $2^+_1$ energy   to 726 keV. This depressed energy is not only interesting from the point of view of nuclear structure \cite{sherrill}, it should also have an important implication for the r-process scenario \cite{fission09}.

Large basis untruncated shell model (SM) calculations have been done in the  valence space consisting of $\pi(1g_{7/2}$,$2d_{5/2}$, $2d_{3/2}$, $3s_{1/2}$, $1h_{11/2}$) and $\nu(1h_{9/2}$, $2f_{7/2}$, $2f_{5/2}$, $3p_{3/2}$,$3p_{1/2}$, $1i_{13/2}$) orbitals above the $^{132}Sn$ core using both realistic CWG \cite{karta:1,sah:1,brwn:2} and empirical SMPN \cite{sah:1,sah:2} (1+2) - body Hamiltonians.  The Hamiltonians  have the same set of single-particle energies  \cite{sah:1} of the valence orbitals but different sets of two-body interaction matrix elements (tbmes).  It should be noted that the neutron-neutron part of the CWG tbmes is the same as that used by Kartamyshev {\it et al.} \cite{karta:1}. The shell model codes OXBASH and NUSHELL@MSU have been used \cite{oxb:1}. The two theoretical results differ dramatically for $^{136,138}Sn$ \cite{sah:1}. The realistic interaction CWG predicts nearly constant energies of $2^+_1$ states for the even-even Sn isotopes above the doubly magic $^{132}Sn$ core, normally expected for singly-magic nuclei. On the other hand, the empirical interaction SMPN predicts decreasing E($2^+_1$)  energies with increasing neutron number. The calculated energies with SMPN fit in the systematics \cite{sah:1} for the experimental E($2^+_1$) energies of their isotones with Z$>$50. They also agree with the trend shown by the Casten - Sherrill systematics for  E($2^+_1$) energy differences of Sn and Te isotopes having same neutron number \cite{sah:1,sherrill,fission09}. It has been shown in Ref.\cite{sah:1} that this non-constancy of E($2^+_1$) in Sn isotopes above $^{132}Sn$ is a strong possibility. The prediction for dramatic decrease of the E(2$^+_1$)  of neutron-rich Sn  with increasing neutron number for N=84-88 using SMPN interaction was considered \cite{sah:1} to be an effect showing weakening of the Z=50 shell gap. But the new result, which we report in this letter, on $^{140}Sn$, its high $2^+_1$ energy and its comparison with examples from other neutron - rich domains clearly show that N=84-88 spectra with SMPN show the effect of gradual filling up of $\nu(2f_{7/2})$  orbital which finally culminates in a new shell closure at N=90. We show that the realistic CWG predicts similar results, that is decreasing $2^+_1$ energies and a shell closure at $^{140}Sn$ if three body effects are included in it.

\begin{figure}[t]
\vspace{4.2cm}
\includegraphics{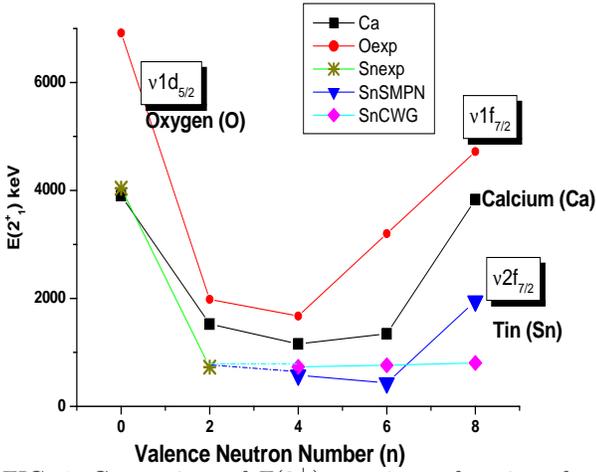}
\vspace{1cm}
\caption {\label{fig1}Comparison of $E(2^+_1)$ energies as function of valence neutron number for oxygen(O),  calcium (Ca) and Sn isotopes.}
\end{figure}

The results for the E($2^+_1$) energies of isotopes of Sn for A=134-140 have been shown in Fig.1 as a function of  valence neutrons above $^{132}Sn$ core. The experimental energies for $^{132,134}Sn$ are shown in the figure. The predicted energies using  CWG and  SMPN interactions are compared. With CWG interaction, as mentioned above, the $0^+_1$ - $ 2^+_1$ spacing remains nearly constant at around 750 keV for $^{136-142}Sn$, except for a small increase at $^{140}Sn$ \cite{karta:1} due to the filling of the  $\nu (2f_{7/2})$ single-particle orbit. It has been identified \cite{karta:1} to be a weak shell closure at $^{140}Sn$. The seniority conserved character of the low lying spectra \cite{casten} observed for Sn isotopes below $^{132}Sn$ is also preserved for isotopes above N=82 shell closure. The nearly constant E($2^+_1$)  value at around 1.2 MeV for $52 \leq N \leq 80$ reduces to around 750 keV above N=82. The same figure also contains variation of the experimental  E($2^+_1$) energies  of even $_{20}Ca$ isotopes from A=40 to 48 as a function of  valence neutrons above $^{40}Ca$ core. Similarly, the variation of the experimental  E($2^+_1$) energies  of even $_{8}O$ isotopes from A=16 to 24 as a function of  valence neutron numbers above $^{16}O$ core are also shown. The variations of experimental E($2^+_1$) with the valence neutron number for two different mass regions and shells, show striking similarity with the theoretical predictions with SMPN in the $^{132}Sn$ region.   The gradual filling of the $\nu (2f_{7/2})$ orbital by neutrons is very distinctively shown by the variation of E($2^+_1$) from $^{134-140}Sn$.  The E($2^+_1$) for $^{140}Sn$ is 1949 keV showing a sudden increase for N=90, indicating a closed shell structure for $^{140}Sn$.
The trend is very similar to that observed for  neutron - rich isotopes of Ca while filling up the $\nu (1f_{7/2})$  orbital and that shown by neutron - rich oxygen isotopes while filling up the $\nu (1d_{5/2})$ single particle orbital (Fig.1).

In order to put forward  further evidence and to understand the shell closure at $^{140}Sn$ more precisely, the effective single-particle energies (ESPE) \cite{otsuka:1} for the  neutron orbitals for the two Hamiltonians have been compared. The ESPE is defined as bare single particle energy (spe) added with the monopole part of the diagonal two body matrix elements (tbme). The bare  spe is originated from the interaction  of a valence nucleon with the doubly closed core. The monopole interaction contribution is the $(2J + 1)$ weighted average of the diagonal tbme, which arises from the interaction of a valence nucleon with the other valence nucleons. The ESPE for the configurations $\nu (2f_{7/2})^n$ in $^{132-140}Sn$ with valence neutron number $n$ varying from 0 to 8 are shown in Fig. 2. For both SMPN and CWG, the energy gap between $\nu (2f_{7/2})$ and $\nu (3p_{3/2})$ single particle orbitals is 854 keV for $^{132}Sn$ core.  But the gap between the corresponding ESPEs increases to 2.246 MeV at N=90 with SMPN.  This gap is sufficient  to  make $^{140}Sn$  a doubly-magic nucleus. For CWG this gap does not show any increase but instead decreases slightly to 826 keV. However shell closure at N=90 with SMPN does not contradict the experimentally observed fact that it is suitable for onset of deformation for nuclei above $^{132}Sn$ ( with Z$\geq$ 54, like Xe, Ba etc.) \cite{nndc,deform,khe:1}. Fig. 3 shows that the proton ESPEs for SMPN  favours the  onset of collectivity at N=90 for Z$\geq$54. This is evidenced by the substantial reduction of the $\pi (1g_{7/2})$ and $\pi (2d_{5/2})$  energy gap  with $\nu (2f_{7/2})^8$. This is 
very similar to  the appearance of the new shell gaps for the oxygen isotopes which disappears at larger Z values
\cite{gor:1}.

\begin{figure}
\vspace{4.5cm}
\includegraphics{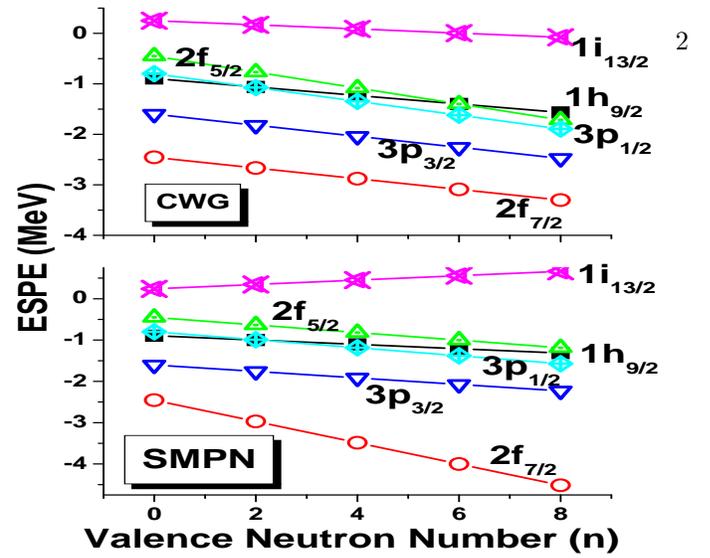}
\vspace{2cm}
\caption{The ESPEs of the neutron single particle orbitals for CWG and SMPN with increasing valence neutron number for Z=50. 
}
\end{figure}

\begin{figure}
\vspace{2.8cm}
\includegraphics{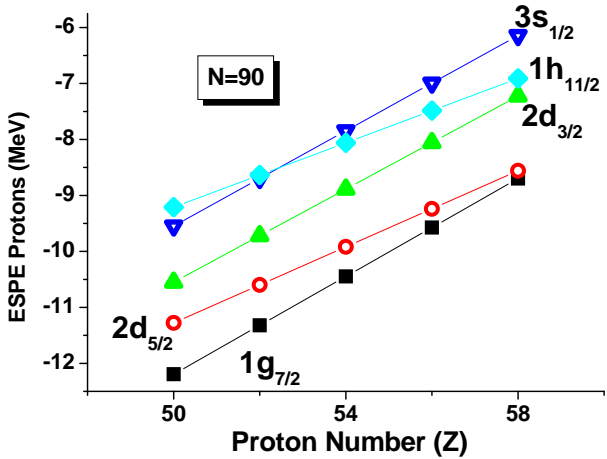}
\vspace{3.cm}
\caption{The ESPEs of the proton single particle orbitals for  SMPN with increasing proton number for N=90. 
}
\end{figure}

 To analyse the origin of this new shell closure, the important physical aspects of both the residual interactions are extracted  by a spin-tensor decomposition \cite {kirson:1} of the two body matrix elements (tbmes). 
The nomenclature for the separated interaction components
has been adopted from Ref. \cite{brow:3}. They are central, antisymmetric spin-orbit (ALS), spin-orbit (LS) and tensor.   Fig.4 shows this decomposition. For SMPN, the central and ALS part for 2$f_{7/2}$- 2$f_{7/2}$
tbmes account for majority of the downward shift of the ESPE of 2$f_{7/2}$with increasing valence neutron number ($n$). The tbmes involving 3$p_{3/2}$ are not modified in SMPN. They have dominant contribution from the central part. The  central parts of 2f$_{7/2}$ and 3p$_{3/2}$ vary with similar slopes for increase in $n$. So the variation in ALS part is primarily responsible for this observed shell gap at N=90.

Even though the neutron-neutron tbmes involving 3$p_{3/2}$ orbital can not be adjusted  at present due to lack of data, it can be safely assumed that the ESPE of this orbital will not have a steeper down-sloping trend than that of 2$f_{7/2}$ closing the presently observed $2f_{7/2}$ - $3p_{3/2}$ gap at N=90. It has been found for Ca isotopes that the monopole contribution of $f_{7/2}$ - $p_{3/2}$ is positive in contrast to the negative values of $f_{7/2}$ - $f_{7/2}$ terms \cite{sorlin:1}.

The  ALS component in the tbmes corresponds to those LS-coupled matrix elements which have
S$\neq$S$^\prime$, i.e., terms non-diagonal in S (spin). Thus these terms do not conserve total spin of the matrix elements \cite{yoshinada:1,brow:4}. But the interactions which are parity conserving and isospin conserving must also conserve the total spin. Bare nucleon-nucleon force contains no ALS term \cite{kirson:1, yoshinada:1, brow:4}. But effective interaction is not simply related to bare nucleon-nucleon force. Core polarisation corrections to the G-matrix give rise to non-zero but small ALS matrix elements. A characteristic feature common to many empirical effective interactions is  strong ALS components in the tbmes \cite{yoshinada:1}. It usually arises from
inadequate constraint by the data. It indicates the important contributions from higher order renormalisation or many body forces to the effective interactions.  In empirical SMPN such many - body effects might have been included in some way through the modification of important tbmes.

\begin{figure}
\vspace{5.8cm}
\includegraphics{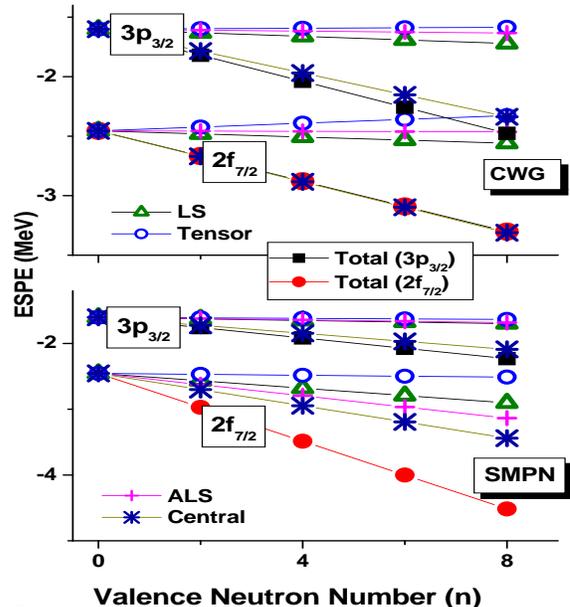}
\vspace{2.5cm}
\caption{The spin - tensor decomposition of neutron  ESPEs   with increasing valence neutron number. 
}
\end{figure}

 At this point it is important to note that SM calculations using two-body realistic interactions derived from the free nucleon-nucleon force fail to reproduce some shell closures \cite{sorlin:1}. It is now rather well established that increase of the $1d_{5/2}$ - $2s_{1/2}$ gap for Z=8 and $1f_{7/2}$ - $2p_{3/2}$ gap for Z=20 (as a function of neutron number), required to explain empirical data are  not obtained in the calculations with these interactions. It has been shown that the three-body forces have to be taken into account to reproduce these shell gaps \cite{zuk:2,zuk:1,sorlin:1,otsuka:2}. Thus many of the previously observed discrepancies are now solved. Otsuka {\it et al.}  \cite{otsuka:2} have proposed a three-body delta-hole mechanism to explain these shell gaps and they have shown that three-body forces are necessary to explain why the doubly-magic $^{24}O$ nucleus  is the heaviest oxygen isotope.  Zuker \cite{zuk:1}  showed earlier that a very simple three-body monopole term can solve practically all the spectroscopic problems in the $p$, $sd$, and $pf$ shells those were hitherto assumed to need drastic revisions of the realistic two-body potentials.

As a next step therefore, in this attempt of analysing the new shell gap, we have incorporated a simple three-body monopole term in CWG as prescribed in Ref.\cite{zuk:1,zuk:2}. We have incorporated corrections in 2$f_{7/2}$-2$f_{7/2}$
and 2$f_{7/2}$-3$p_{3/2}$ tbmes similar to those in KB3 for 1$f_{7/2}$-1$f_{7/2}$ and 1$f_{7/2}$-2$p_{3/2}$ tbmes. 
The correction terms included in the tbmes are $V^{J,T=1}_{ffff}$(CWG3)$=$ $V^{J,T=1}_{ffff}$(CWG)$-$110 keV, for J=0,4 and 6;  $V^{J=2,T=1}_{ffff}$(CWG3)$=$ $V^{J=2,T=1}_{ffff}$(CWG) $-$ 310 keV and  $V^{J,T=1}_{frfr}$(CWG3)$=$ $V^{J,T=1}_{frfr}$(CWG)+300 keV for J=2, 3, 4  and 5. Here f stands for $2f_{7/2}$  and r stands for $3p_{3/2}$.  The effective single particle energies after this correction are plotted in Fig. 5. A stronger $2f_{7/2}-3p_{3/2}$
shell gap now appears with CWG3 compared to that with SMPN. 

With this new interaction, CWG3, the binding energy of $^{134}Sn$ and $E(2^+_1$) energies of 
$^{134-140}Sn$  are -6.307, 0.627, 0.562, 0.521 and 2.532 MeV, respectively. This trend is very similar to that obtained with SMPN, with an indication of a stronger shell gap at N=90. Comparison of the wave functions \cite{sah:4} 
of the $0^+_{g.s.}$ and $2^+_1$ states with CWG and SMPN showed that whereas CWG favours large configuration mixing conserving seniority as far as possible, the SMPN on the other hand favours purer structure of the low-lying states, characteristics of $\nu (2f_{7/2})^8$ and $\nu (2f_{7/2})^7(3p_{3/2})$ multiplets. For CWG3, the wave function composition for the $0^+_{g.s}$ is (81.3\%) from the $\nu (2f_{7/2})^8$ partition, similar to SMPN. But due to
overestimation of the up-sloping trend of $\nu(3p_{3/2})$ ESPE (Fig.4), for $2^+_1$ state, 
38.7\% originates  from the  $\nu (2f_{7/2})^6(1h_{9/2})^2$ and 12\% from  $\nu (2f_{7/2})^6(2f_{5/2})^2$. 

\begin{figure}
\vspace{3.cm}
\includegraphics{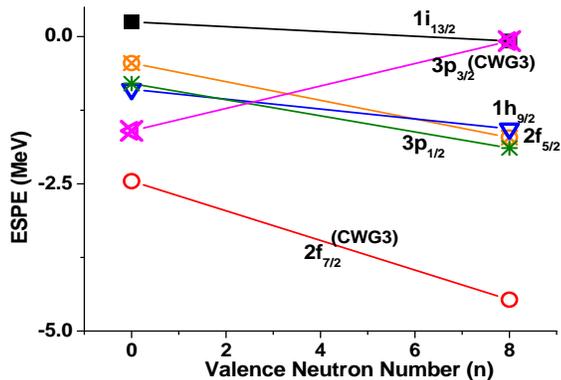}
\vspace{1.2cm}
\caption{The neutron  ESPEs  for CWG3 with increasing valence neutron number. (see text for details). 
}
\end{figure}

In conclusion we find that comparison with the systematics of $2^+_1$ energies of other n-rich domains and the
spin-tensor decomposition of the two interactions establish the new shell closure at $^{140}Sn$.
The ALS term incorporates in it the contributions of many body forces in the empirical interaction. A large contribution of this term in the ESPE of 2f$_{7/2}$ in empirical interaction SMPN has been observed. It is found to be responsible for the gap observed in SMPN results. The CWG indicates a weak shell closure at $^{140}Sn$. A simple three-body monopole term  has been included in CWG to get CWG3. The new  CWG3 predictions showed good agreement with that from SMPN, indicating a stronger shell gap at N=90 for Sn isotopes as well as decreasing $2^+_1$ energies for $^{134-138}Sn$. This also shows,  similar to that in $sd$  and $fp$ shells, three body effect plays an important role for shell evolution in neutron rich Sn isotopes above  $^{132}Sn$. The anomalously depressed $2^+_1$ states in Sn isotopes having  N=84-88, and the new magic number  for N=90, might have interesting consequences for the r - process nucleosynthesis.  

The  authors  thank  Prof.  Waldek  Urban  for   stimulating discussions on the issues in this mass region. Special thanks are due to Prof. B.A. Brown for his help in providing us  the  OXBASH (Windows Version) and the Nushell@MSU codes.

\end{document}